\documentclass[twocolumn]{emulateapj}

\usepackage{amsmath}
\usepackage[title]{appendix}

\shorttitle{Initial-Final Mass Relation and Rotation}
\shortauthors{Cummings et~al.}

\begin{document}

\title{A Novel Approach to Constrain Rotational Mixing \& Convective-Core Overshoot\\ in Stars Using 
the Initial-Final Mass Relation}

\author{Jeffrey D. Cummings\altaffilmark{1}, Jason S. Kalirai\altaffilmark{2,3}, Jieun Choi\altaffilmark{4}, C. Georgy\altaffilmark{5}, \\P.-E. Tremblay\altaffilmark{6}, AND Enrico Ramirez-Ruiz\altaffilmark{7}}
\affil{}

\altaffiltext{1}{Center for Astrophysical Sciences, Johns Hopkins University,
3400 N. Charles Street, Baltimore, MD 21218, USA; jcummi19@jhu.edu}
\altaffiltext{2}{Johns Hopkins University Applied Physics Laboratory, 11101 Johns Hopkins Road, Laurel, MD 20723, USA; Jason.Kalirai@jhuapl.edu}
\altaffiltext{3}{Space Telescope Science Institute, 3700 San Martin Drive, Baltimore, MD 21218, USA}
\altaffiltext{4}{Harvard-Smithsonian Center for Astrophysics, Cambridge, MA
02138, USA; jieun.choi@cfa.harvard.edu} 
\altaffiltext{5}{Geneva Observatory, University of Geneva, Maillettes 51, 1290 Sauverny, Switzerland; 
cyril.georgy@unige.ch} 
\altaffiltext{6}{Department of Physics, University of Warwick, Coventry CV4 7AL, UK; 
P-E.Tremblay@warwick.ac.uk} 
\altaffiltext{7}{Department of Astronomy and Astrophysics, University of California,
Santa Cruz, CA 95064; enrico@ucolick.org} 

\begin{abstract}
The semi-empirical initial-final mass relation (IFMR) connects spectroscopically analyzed white dwarfs in star 
clusters to the initial masses of the stars that formed them.  Most current stellar evolution models, however, 
predict that stars will evolve to white dwarfs $\sim$0.1 M$_\odot$ less massive than that found in the IFMR.   
We first look at how varying theoretical mass-loss rates, third dredge-up efficiencies, and convective-core 
overshoot may help explain the differences between models and observations.  These parameters play an important 
role at the lowest masses (M$_{\rm initial}$ $<$ 3 M$_\odot$).  At higher masses, only convective-core overshoot 
meaningfully affects white dwarf mass, but alone it likely cannot explain the observed white dwarf masses nor why 
the IFMR scatter is larger than observational errors predict.  These higher masses, however, are also where 
rotational mixing in main sequence stars begins to create more massive cores, and hence more massive white dwarfs.  
This rotational mixing also extends a star's lifetime, making faster rotating progenitors appear like less massive stars
in their semi-empirical age analysis.  Applying the observed range of young B-dwarf rotations to the MIST or SYCLIST 
rotational models demonstrates a marked improvement in reproducing both the observed IFMR data and its scatter.  The 
incorporation of both rotation and efficient convective-core overshoot significantly improves the match with observations.  
This work shows that the IFMR provides a valuable observational constraint on how rotation and convective-core overshoot affect 
the core evolution of a star.
\end{abstract}

\section{Introduction}

How mass loss, third dredge-up (hereafter 3DUP), convective-core overshoot (hereafter CCO), and rotation affect 
a star and its core evolution are long-standing challenges to model.  Rotation is one of the most complex 
processes involved (e.g., Maeder \& Meynet 2000, Langer 2012).  Rapid rotation makes stars non-spherical, 
introducing strong changes to a star's observed characteristics, and rotation also introduces multiple types of 
interior mixing processes that non-, or slowly, rotating stars do not experience.  This mixing can significantly 
affect how a star evolves by bringing fresh hydrogen into its core, prolonging the hydrogen burning phase 
and increasing the total mass of the hydrogen-exhausted core (e.g., Talon et~al.\ 1997).

The MIST (Dotter 2016, Choi et~al.\ 2016, 2017) and SYCLIST (Georgy et~al.\ 2013, 2014) models consider the 
effects of rotation throughout all, if not nearly all, stages of stellar evolution.  Both of these look at a broad 
range of initial-rotation rates from non-rotating to $\sim$0.80 of critical rotation velocity and consider rotation's 
effect on a star's observed characteristics, their core evolution, and their lifetime.  There exist several types 
of observations indicative of these effects: 1) Asteroseismology of rapidly rotating Be stars finds that they have 
abnormally high-mass cores (Neiner et~al.\ 2012).  2) Observations of young clusters suggest that faster-rotating
stars evolving more slowly is likely an important component of broad cluster turnoffs (e.g., Brandt et~al.\ 2015, 
Niederhofer et~al.\ 2015, D'Antona et~al.\ 2015).  3) Rotational mixing in stars can help explain surface-abundance trends of 
various elements like carbon, nitrogen, and the light elements (e.g., Hunter et~al.\ 2009, Proffitt et~al.\ 2016,
Cummings et~al.\ 2017).  

The consequences of rotation, in general, remain qualitatively consistent between various rotational models, but 
large differences remain in their predicted magnitudes.  This is due to the complexity of rotation and 
rotational mixing, the difficulty of acquiring several of these observational tracers, that certain observations 
only trace mixing in the envelope, and that additional processes may affect these tracers.

\begin{figure*}[!ht]
\begin{center}
\includegraphics[trim=0.5cm 11.3cm 0.5cm 3.2cm, width=1.03\textwidth]{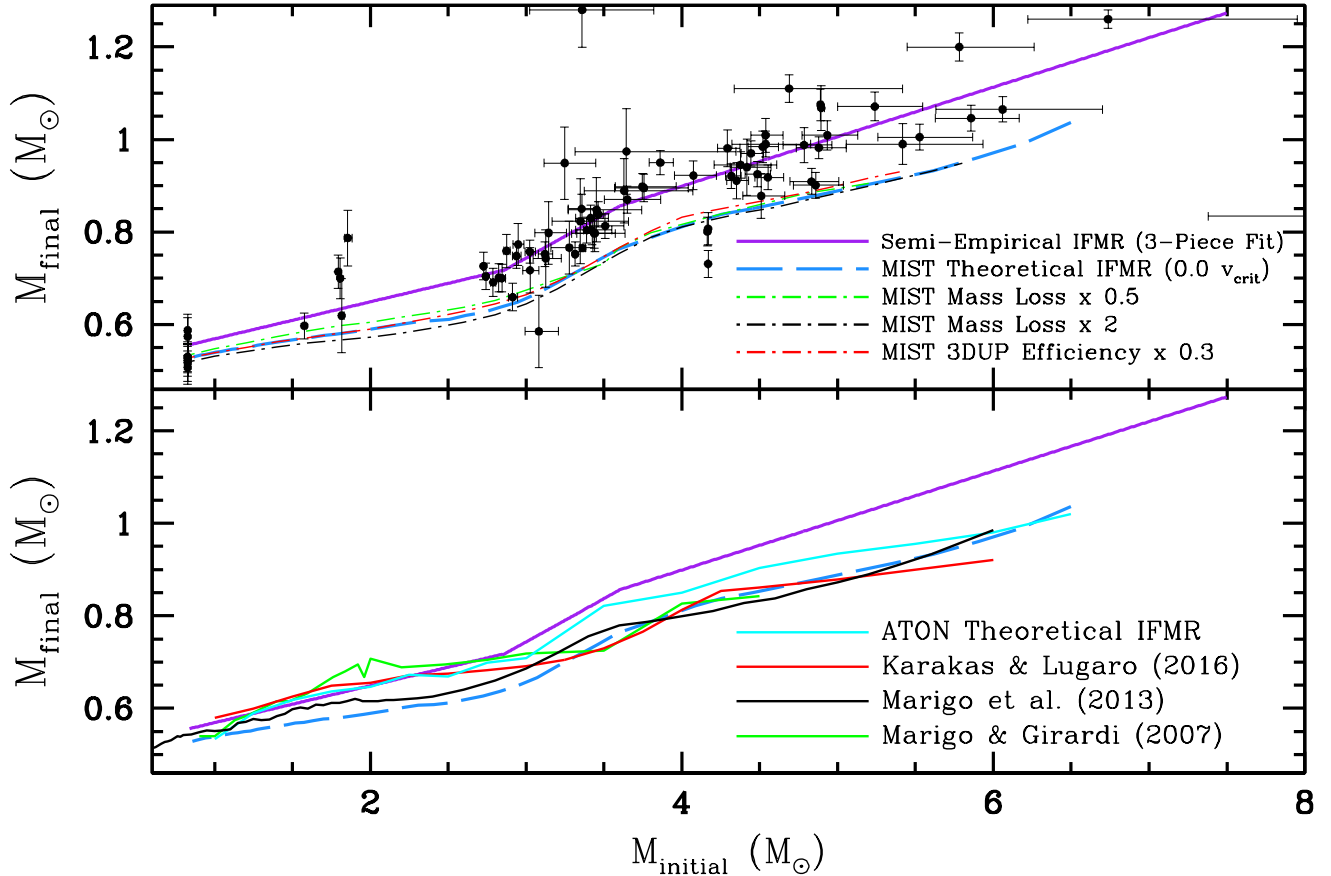}
\end{center}
\vspace{-0.4cm}
\caption{The upper panel compares the semi-empirical IFMR from Paper I to the non-rotating MIST theoretical 
IFMR.  We illustrate that the IFMR has minor sensitivity to mass-loss rates and 3DUP efficiencies at the 
lowest masses ($M_{\rm initial}$ $<$ 3 M$_\odot$) and even less sensitivity at higher masses.  The lower 
panel illustrates this further with independent theoretical IFMRs at solar metallicity, which show important 
differences at lower masses but are primarily consistent at higher masses.  Only the ATON model with increased 
CCO predicts distinctly higher M$_{\rm final}$, which can partially explain the offset from observations but 
not its intrinsic scatter.}
\end{figure*}

Additional observational constraints on rotation would be valuable, and ideally ones that are sensitive to mixing 
at the cores of stars rather than only their outer envelopes.  This letter analyzes how white dwarfs, the hot and 
exposed remnants of these stellar cores, can observationally constrain rotational mixing's and CCO's effect on core 
evolution.

This letter's structure is as follows: In Section 2 we discuss the initial-final mass relation of stars and the disagreements 
between observations and the available non-rotating models.  We also discuss stellar evolution's sensitivity to 
mass-loss rates, 3DUP, and CCO.  In Section 3 we demonstrate how the IFMR is sensitive to progenitor rotation rates
and can constrain rotational mixing in the cores of stars.  In Section 4 we summarize the results.

\vspace{0.5cm}
\section{The Initial-Final Mass Relation}

The initial-final mass relation (hereafter IFMR) compares a star's initial main sequence mass to its final mass 
after it evolves to a white dwarf (hereafter WD) and has long been a tool to constrain stellar evolution
models (e.g., Koester \& Weidemann 1980).  Semi-empirical IFMRs are commonly based on spectroscopic analysis of WDs 
in star clusters (e.g., Weidemann 2000, Kalirai et~al.\ 2008, Cummings et~al.\ 2016a, 2018; hereafter Paper I).  Spectroscopic 
fitting of the Balmer lines in hydrogen-rich WDs measures both their T$_{\rm eff}$ and log g.  Applying 
these parameters to WD cooling models gives a WD's current mass, luminosity, and cooling age, 
which is the time since it has left the tip of the asymptotic giant branch (hereafter AGB).  

The method of deriving the IFMR first compares each WD's spectroscopically derived photometry to its observed 
apparent photometry, which tests its cluster membership and single star status.  The second step directly 
compares a WD member's cooling age to its cluster's total age.  This gives the evolutionary lifetime of its 
progenitor, and with application to models, ideally the same evolutionary models used to derive the cluster's age, 
this yields the initial mass of each WD's progenitor (Cummings \& Kalirai 2018, Paper I).

The upper-panel of Figure\,\,1 presents the semi-empirical IFMR from Paper I, which analyzed 80 WDs and 
used the MIST non-rotating models from Choi et~al.\ (2016) to both derive cluster ages and infer M$_{\rm initial}$ 
(see Cummings \& Kalirai 2018 for discussion of cluster age analysis).  In solid purple, we show a linear 3-piece 
continuous fit, and in dashed-blue we show the theoretical non-rotating IFMR from MIST, which predicts 
progenitors will form WDs $\sim$0.1 M$_\odot$ less massive than observations.  Additionally, even though 
this IFMR's scatter has significantly decreased relative to previous semi-empirical IFMRs, the data's
M$_{\rm initial}$ and M$_{\rm final}$ errors have a moderate positive correlation (Cummings et~al.\ 2016b), 
which makes the scatter at higher masses ($>$ 3 M$_\odot$) significantly larger than these observational 
and cluster-age errors can explain.  Therefore, an intrinsic IFMR scatter may also be needed to explain 
observations.

In the upper-panel of Figure\,\,1, we consider mass-loss rates in the MIST evolutionary models by increasing 
(dashed-black) or decreasing (dashed-green) the applied mass-loss rate by a factor of 2 at \textit{all} stages 
of evolution.  This shows that while larger variations in mass-loss rates may play some role in lower-mass 
progenitors ($<$ 3 M$_\odot$), at higher masses (3 to 6.5 M$_\odot$) the WDs are increasingly
insensitive to progenitor mass-loss rates.  

This minor to very weak sensitivity to mass-loss rates at all phases is first because before 
the AGB, the standard mass-loss rates are negligible compared to the total mass of a star; hence, even large 
variations of mass-loss rates during these phases play little role in the core-evolution of a star.  Second, during 
the AGB the mass-loss rates increase significantly, and in lower-mass stars, the core-masses grow as much as 30\%\, 
in the thermally-pulsing AGB phase (e.g., Kalirai et~al.\ 2014).  Therefore, at these lower masses, an increase in 
the AGB mass-loss rate can cut this rapid core-mass growth short.  At higher-masses, though, the 
thermally-pulsing AGB phase is rapid and little change in core mass occurs during this phase (e.g., 
Marigo et~al.\ 2013).  Moderately changing mass-loss rates or even introducing stochastic fluctuations in mass 
loss will not significantly affect the resulting WDs at higher masses (Doherty et~al.\ 2014).

In the upper-panel of Figure\,\,1, we also consider 3DUP efficiency, which plays a direct role in regulating 
the core-mass evolution during the thermally-pulsing AGB phase (Kalirai et~al.\ 2014).  More efficient 3DUP 
limits the core-mass growth during this phase while a lower efficiency lets the core grow with weaker
reduction episodes.   
The dashed red line represents the theoretical MIST IFMR with 3DUP efficiency decreased to 30\% of the standard.
This shows it can affect WD masses, but even where it is the most important ($\sim$4 M$_\odot$) the resulting 
WDs are only increased in mass by $\sim$0.02 M$_\odot$.

We additionally note that even though we are changing model parameters, these primarily only affect the relatively 
short AGB phase and have a negligible effect on theoretical evolutionary timescales.  Hence, the semi-empirical
M$_{\rm initial}$ are not affected.  

These comparisons show that the uncertainties remaining in mass-loss rates and 3DUP may play a role in 
better matching WD masses, but at higher masses ($>$ 3 M$_\odot$) these uncertainties alone cannot explain 
observations.  In the lower-panel of Figure\,\,1 this is further illustrated by comparing multiple independent 
theoretical IFMRs.  These IFMRs have important differences at lower masses, but virtually all 
models converge below observations at higher masses.  The only theoretical IFMR with a meaningful difference is the ATON IFMR of 
Ventura et~al.\ (2018), which adopts very similar physics to Karakas \& Lugaro (2016) but with more CCO that 
leads to increased WD masses.  The ATON model CCO also adopts the exponentially diffusive overshoot 
prescription, as do the MIST models, but again using a higher efficiency.  This can partially explain the offset 
from observations but not its intrinsic scatter.  Therefore, a non-standard process may be needed.

\section{The Effects of Rotation on the IFMR}

Stellar rotation induces mixing that drives additional hydrogen fuel into the core of a star, which has two 
consequences for the IFMR.  First, faster-rotating progenitors evolve more massive cores, producing more 
massive WDs (Dominguez et~al.\ 1996).  In Figure\,\,2 we show for the MIST models the quantitative effects of 
rotation on the IFMR for stars at M$_{\rm initial}$ of 4, 5, and 6 M$_\odot$.  Each color represents a different 
initial rotation, with open circles representing the direct effect of how rotation creates higher-mass cores, 
leading to higher-mass WDs.  This is a true shift of the IFMR to higher masses, 
creating an intrinsic spread in the IFMR due to the broad range of rotation rates observed in B dwarfs (e.g., 
Huang et~al.\ 2010).  

Second, faster rotating progenitors evolve more slowly.  Therefore, in the IFMR analysis, a fast rotating progenitor 
will appear identical to a lower-mass slowly-rotating progenitor.  This does not directly affect the true IFMR, 
but because non-rotating models have been adopted to infer M$_{\rm initial}$, it introduces an offset in the 
determination of a WD's M$_{\rm initial}$.  In Figure\,\,2 for each M$_{\rm initial}$ of 4, 5, and 6 M$_{\odot}$ 
we show with x's how much a given initial-rotation rate will cause the inferred M$_{\rm initial}$ to be underestimated.  
The solid circles in Figure\,\,2 represent this effect in combination with the intrinsically higher-mass WDs.   
In the semi-empirical derivation of the IFMR, both of these factors combine to introduce a large scatter by
shifting WDs formed by faster rotating progenitors to systematically higher masses and lower inferred 
M$_{\rm initial}$.

\begin{figure}[!ht]
\begin{center}
\includegraphics[trim=0.5cm 5.1cm 0.5cm 3.2cm, width=0.5\textwidth]{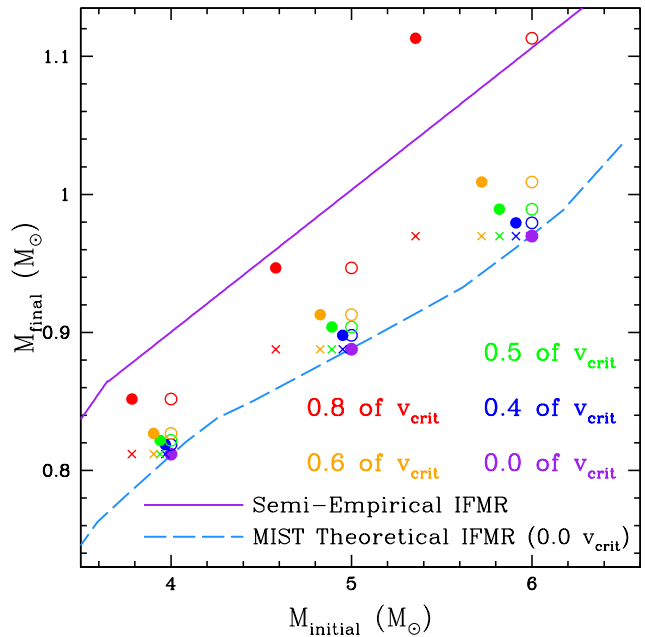}
\end{center}
\vspace{-0.4cm}
\caption{The effects of rotation in the MIST models at M$_{\rm initial}$ of 4, 5, and 6 M$_\odot$ for WD
masses only (open circle), for systematic effects on the inference of M$_{\rm initial}$ from evolutionary 
timescales (x), and for these effects combined (solid circle).  This can help to reproduce both the observed 
WD's higher masses and their scatter.}
\end{figure}

\begin{figure*}[!ht]
\begin{center}
\includegraphics[trim=0.5cm 5.5cm 0.5cm 3.2cm, width=1.03\textwidth]{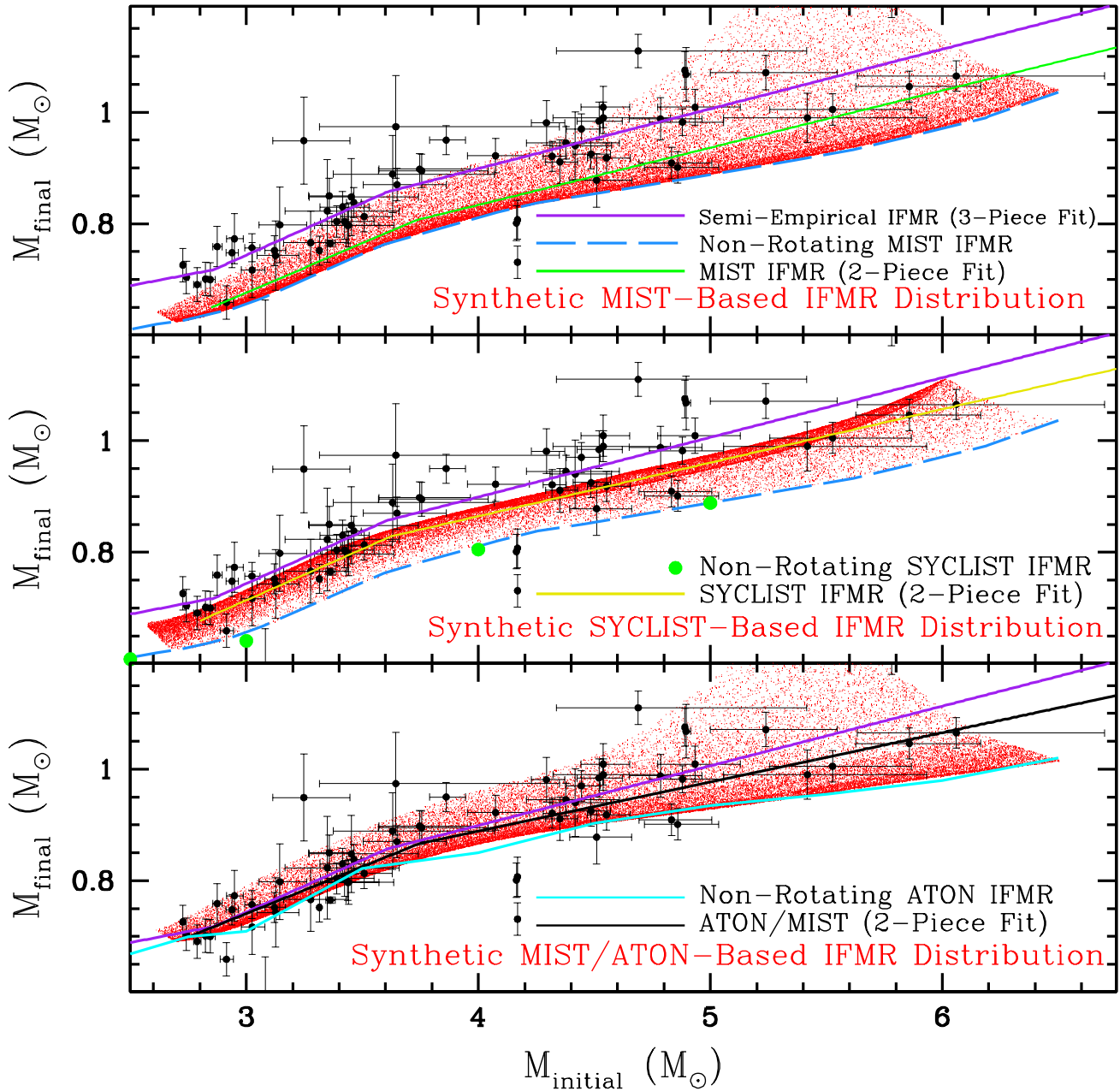}
\end{center}
\vspace{-0.4cm}
\caption{Application of the Huang et~al.\ (2010) rotational distribution to the MIST rotational models (upper 
panel), to the SYCLIST rotational models (middel panel), and to the MIST rotational models applied to a smoothed 
ATON IFMR (lower panel).  2-piece fits are shown for the MIST distribution (green), the SYCLIST distribution 
(gold), and the MIST/ATON distribution (black).  In the middle panel the consistency of the non-rotating SYCLIST 
IFMR (green data) and the non-rotating MIST IFMR (dashed blue) are shown.}
\end{figure*}

\subsection{Synthetic Initial-Final Mass Relation}

\begin{figure}[!ht]
\begin{center}
\includegraphics[trim=0.5cm 5.1cm 0.5cm 3.2cm, width=0.5\textwidth]{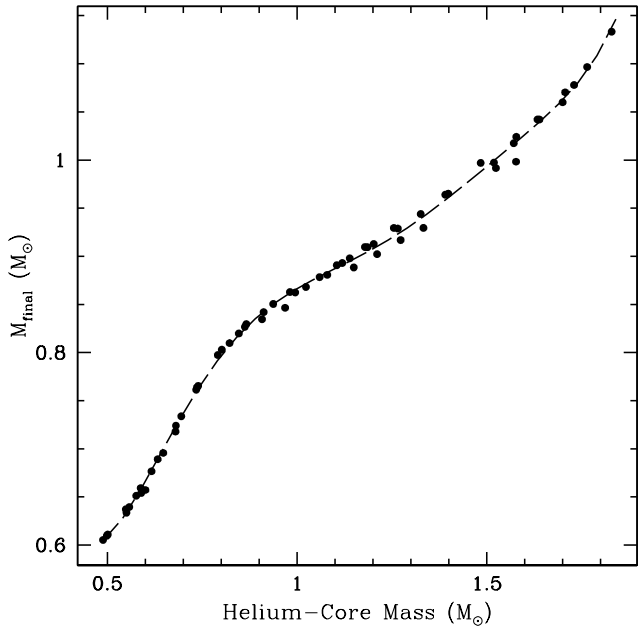}
\end{center}
\vspace{-0.4cm}
\caption{MIST model relation between helium-core mass at the beginning of the AGB and final WD
mass.  This relation is insensitive to rotation and has been applied to SYCLIST helium-core masses to further
study the IFMR.}
\end{figure}

\begin{figure*}[!ht]
\begin{center}
\includegraphics[trim=0.5cm 5.6cm 0.5cm 8.5cm, width=1.0\textwidth]{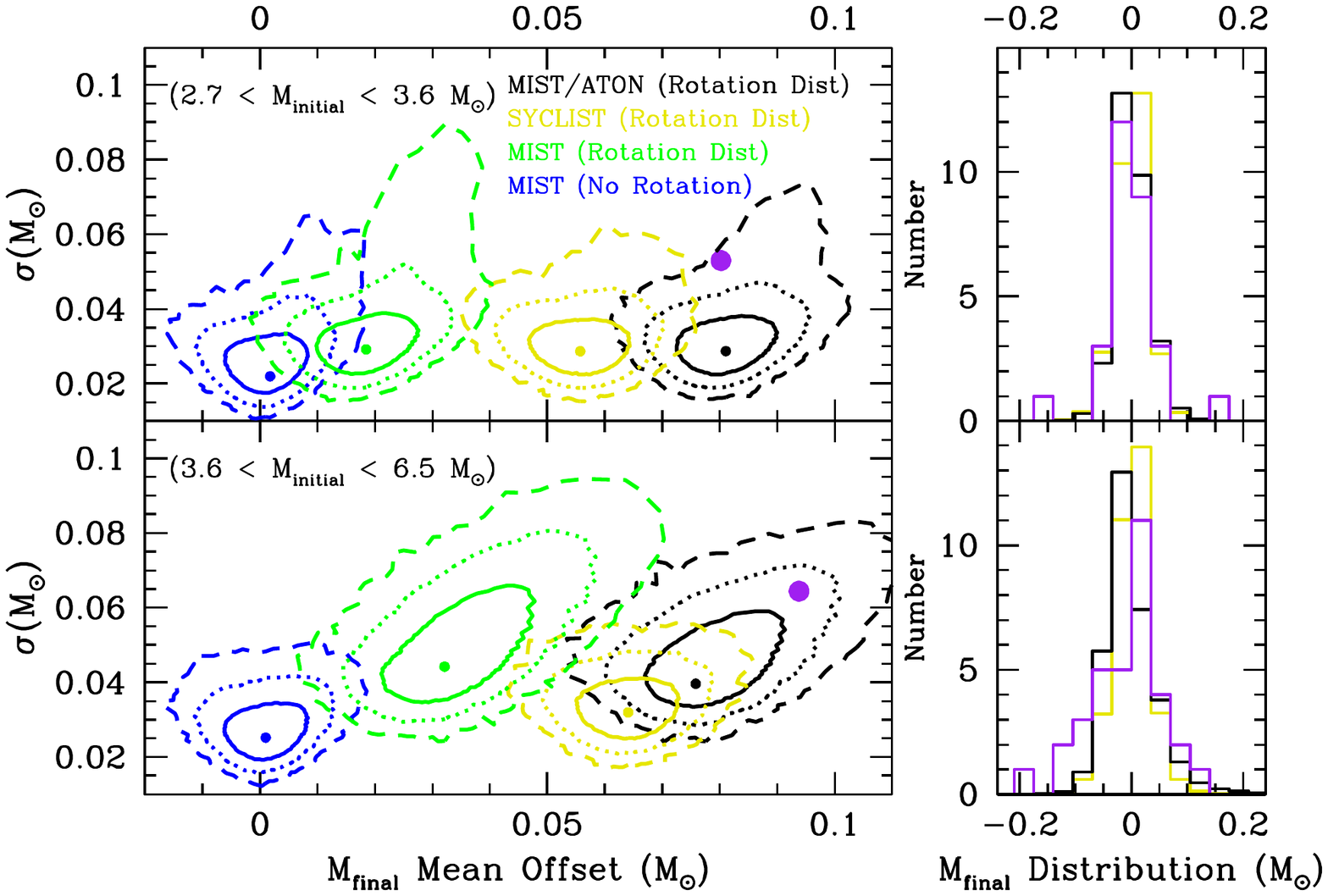}
\end{center}
\vspace{-0.4cm}
\caption{Observational errors and an IMF are applied to the synthetic IFMRs.  Monte Carlo analysis of these 
distributions are performed by drawing the same number of synthetic WDs that have been observed at intermediate 
and high masses (upper and lower panels, respectively).  2D contours representing the distribution peaks and 1 (solid), 
2 (dotted), and 3 (dashed) $\sigma$ show that with MIST rotation applied to the ATON IFMR (black), it can match observations (purple), 
but that the observed scatter is only partially explained.  The right panels illustrate the M$_{\rm final}$ 
residual distributions relative to observations (purple).}
\end{figure*}

While methods exist to measure WD rotations (e.g., Koester et~al.\ 1998, Kilic et~al.\ 2015), these
observations are challenging.  Further, it is difficult to use WD rotations to infer information about their 
progenitor's rotational histories (Kawaler 2015, Hermes et~al.\ 2017).  Therefore, with this large IFMR sample, 
we instead consider the effects of rotation statistically.  Huang et~al.\ (2010) observed young (high log g) 
B dwarfs (ranging from approximately 2 to 10 M$_\odot$) in the field and open clusters.  They found a broad 
distribution of rotations spanning from several percent of $v_{crit}$ to $\sim$0.95 of 
$v_{crit}$, with the most common rotation for young B dwarfs being $\sim$0.49 of $v_{crit}$.  We note that their 
rotational distribution appears to be moderately sensitive to mass, with the higher masses still having a broad 
range of rotations but preferentially rotating at a lower percentage of their $v_{crit}$.  The models of Rosen 
et~al.\ (2012), however, find this results from more rapid angular momentum loss at higher masses and is not a 
result of slower initial rotation.  Therefore, we will uniformly apply the total B-dwarf 
rotational distribution from Huang et~al.\ (2010) for all masses.

In the upper panel of Figure\,\,3, we draw 40,000 synthetic stars in red (based on MIST rotational models) that are uniformly 
distributed in M$_{\rm initial}$ and follow the Huang et~al.\ (2010) rotation distribution.  This synthetic IFMR 
includes both the intrinsic increase in WD masses and, for the purpose of comparison to observations, the 
corresponding systematic estimate of a lower M$_{\rm initial}$.  This resulting scatter comes from the 
broad distribution of progenitor rotations, but note that the distribution remains concentrated at the lower 
envelope.  This is because the MIST rotational models adopt a lower rotational-mixing efficiency and require above 
average rotation rates before the mixing becomes important.  Therefore, this only systematically shifts the mean 
WD trend upward by 0.02 M$_\odot$ (M$_{\rm initial}$ of 3 to 3.6 M$_\odot$) and 0.04 M$_\odot$ (M$_{\rm initial}$ 
of 3.6 to 6 M$_\odot$).  This alone is not enough to match observations. 

The magnitude of rotational effects in models still remains poorly constrained.  Therefore, we also consider the 
SYCLIST rotational models, which adopt more efficient rotational mixing.  The SYCLIST models do not fully evolve 
to the WD cooling sequence, but at this intermediate-mass range, they all do evolve at least to the end of central helium 
burning (the beginning of the AGB).  The phases before the AGB are where the direct effects of rotation 
are important.  Stars have lost enough of their angular momentum before they reach the short AGB phase that 
any remaining variations in angular momentum no longer play a major role.  However, more rapidly 
rotating stars at early stages do evolve more massive cores, and based on the MIST models the remaining core 
evolution during the AGB is predominantly sensitive to the helium-core mass, rather than the total mass, 
at the beginning of the AGB; differences in initial rotation and remaining envelope mass play no major role.  
Figure\,\,4 illustrates the consistent relation between these two masses from MIST across all rotation rates.

We apply Figure\,\,4's relation to SYCLIST helium-core masses at the beginning of the AGB to quantify the SYCLIST
IFMR's sensitivity to rotation.  In the middle-panel of Figure\,\,3, the resulting IFMR for the non-rotating SYCLIST 
models are shown as green data points at M$_{\rm initial}$ of 2.5, 3, 4, and 5 M$_\odot$ and are comparable to 
that from MIST (dashed-blue).  Applying the SYCLIST model's rotational sensitivity to the consistent, but more 
complete, non-rotating MIST IFMR produces the synthetic IFMR distribution shown.  Across this M$_{\rm initial}$ 
range the SYCLIST models give a more consistent IFMR spread in M$_{\rm final}$, and its density distribution is 
notably different with a weakly populated lower envelope and a densely populated upper envelope.  This results 
from more efficient rotational mixing requiring little rotation to affect evolution, followed by rotational 
mixing reaching a saturation point at higher rotation, producing a concentration in the IFMR.  The SYCLIST 
synthetic IFMR shifts the mean trend upward by $\sim$0.06 M$_\odot$.  

These two synthetic IFMRs still fall short of observations, but the ATON IFMR with increased CCO further increases 
core mixing.  In the lower-panel of Figure\,\,3, we apply the MIST rotational models to a smoothed ATON non-rotating 
IFMR.  This produces a much stronger consistency with observations, with a mean trend shifted 
upward by $\sim$0.08 M$_\odot$ relative to the non-rotating MIST IFMR.  However, unlike rotational-mixing, CCO 
does not increase scatter.  Additionally, note that increased CCO also extends a star's lifetime, but unlike 
rotation it uniformly affects all stars of a given mass and systematically affects derived cluster age, so it does 
not significantly affect the inference of M$_{\rm initial}$ (see Paper I).

\subsection{Monte Carlo Analysis}

Monte Carlo analysis provides a more powerful comparison between a synthetic IFMR and observations.  For 
each synthetic IFMR model from Figure\,\,3, we first apply an M$_{\rm initial}$ distribution based directly on 
the data (power law of exponent=$-$2.45) and generate 10 million WDs.  Second, we apply a distribution of 
observational errors based directly on the data.  Third, to match the observed statistics, we match the observed 
numbers by drawing 29 intermediate-mass synthetic WDs (2.7 $<$ M$_{\rm initial}$ $<$ 3.6 M$_\odot$) and 34 
higher-mass synthetic WDs (3.6 $<$ M$_{\rm initial}$ $<$ 6.5 M$_\odot$).  By synthetically applying the 
observational errors, numbers, and M$_{\rm initial}$ distribution to each evolutionary and 
rotational model, we can more directly compare to the semi-empirical IFMR trends and scatter.

In Figure\,\,5's left panels, for each synthetic IFMR model, we illustrate on the x-axis 
each intermediate- and high-mass synthetic sample's mean M$_{\rm final}$ offset from the non-rotating 
MIST IFMR.  On the y-axis we illustrate the corresponding $\sigma$ relative to each synthetic IFMR's 
2-piece fit in Figure\,\,3.  We illustrate each distribution's central peak and 1, 2, and 3$\sigma$ contours,
and this shows how these parameters are correlated.  We represent the 
distributions for MIST models with no applied rotation in blue and full rotational models in green,
SYCLIST rotational models in gold, and MIST rotational models applied to the smoothed non-rotating ATON 
IFMR in black.  These are compared to the semi-empirical IFMR in purple.
In the right panels, histograms of the resulting M$_{\rm final}$ scatter distribution shapes show MIST 
rotational models with their concentration on the lower envelope and SYCLIST rotational models 
with their concentration on the upper envelope.

This illustrates the impact of rotation in IFMR analysis and the resulting differences between models. 
Additionally, the MIST/ATON combination can recreate the high-mass observations within 2$\sigma$ and the 
intermediate-mass observations within 3$\sigma$, but the synthetic scatter remains typically smaller.  Additionally, 
the right panels show that the model distribution shapes are comparable to observations, but at higher masses the 
MIST/ATON model's distribution appears inconsistent.

\section{Summary}

The semi-empirical IFMR is a valuable constraint of stellar evolution.  Comparisons to various recent models 
show that at the lowest masses (M$_{\rm initial}$ $<$ 3 M$_\odot$) there are still important differences in predicted
M$_{\rm final}$.  At lower masses these differences predominantly result from how these models handle the 
AGB.  Significant limitations remain in the low-mass data, however, currently making it challenging to 
constrain these models, but appropriate adjustments to AGB parameters alone will likely be able to explain 
low-mass observations.  Additionally, at these lower masses, an intrinsic IFMR scatter is likely not 
needed because the intracluster scatters are consistent with observational errors (Paper I, Williams 
et~al.\ 2018).  Consistent with this, at lower masses the effects of rotation on core evolution are 
not strong enough to produce an intrinsic scatter.

For intermediate and higher masses (3 to 6.5 M$_\odot$), the theoretical IFMRs are less sensitive to the AGB 
model parameters.  In comparison to observations, nearly all models predict WDs masses $\sim$0.1 M$_\odot$ 
below observations.  Realistic changes in mass-loss rates and 3DUP efficiencies can play some role but likely 
cannot explain this difference nor the increased scatter.  The ATON models, however, do show that 
efficient exponentially diffusive CCO is a powerful way to help match, but not fully explain, 
observations.  

All of these models, however, have assumed non-rotating progenitors.  Therefore, we have demonstrated in this letter
the effects that progenitor rotation will have on the IFMR.  Applying the effects of rotation from either the MIST 
or SYCLIST rotational models to a non-rotating IFMR shows that through the broad range 
of B-dwarf rotations, both models predict a broad IFMR scatter but with important differences; the 
MIST models with less efficient rotational-mixing predict a higher 
concentration of stars on the distribution's lower envelope while the SYCLIST models predict a higher concentration 
of stars on the upper envelope.  These result in systematically higher WD masses relative to the non-rotating 
model, with MIST models shifting 0.02 to 0.04 M$_\odot$, SYCLIST models shifting 0.06 M$_\odot$, 
and MIST/ATON models shifting 0.08 M$_\odot$.

Applying observational errors, an IMF, and the Monte Carlo method to these synthetic IFMRs shows that these 
rotational models can also explain most, but likely not all, of the observed scatter being larger than 
observational errors.  Therefore, further refinement of rotational mixing efficiencies may be needed, but 
we note that the effects of environment, field contaminants, or merger remnants could also be the cause of this 
increased observed IFMR scatter.  More cluster white dwarf data are needed to study this further 
because other available methods to constrain the IFMR, e.g., with Gaia (El-Badry et~al.\ 2018), are unable 
to characterize the IFMR scatter.  This letter, however, illustrates that the semi-empirical IFMR is a powerful 
tool for constraining the effects of rotation and CCO on the core evolution of stars. 

\vspace{0.5cm}
Acknowledgments: 
This project was supported by the National Science Foundation through grant AST-1614933 and the European 
Research Council under the European Union's Horizon 2020 research and innovation programme n. 677706 (WD3D).

\end{document}